\documentclass[]{rmaa}
\newcommand{\map}{\hbox{{\sc mappings i}c}}
\newcommand{\ii}{\'{\i}}
\def\ion#1#2{#1$\;${\small\rm\@Roman{#2}}\relax}
\newcommand{\up}{\hbox{$U$}}

\newcommand{\cmc}{\hbox{\,${\rm cm^{-3}}$}}

\newcommand{\gama}{\hbox{$\Gamma_{abs.}$}}
\newcommand{\gamh}{\hbox{$\Gamma_{heat.}$}}
\newcommand{\teff}{\hbox{$T_{eff}$}}
\newcommand{\tsq}{\hbox{$t^2$}}
\newcommand{\teq}{\hbox{$T_{eq}$}}
\newcommand{\tmean}{\hbox{$\bar T_0$}}
\newcommand{\trec}{\hbox{$\bar T_{rec}$}}

\newcommand{\ha}{\hbox{H$\alpha$}}

\newcommand{\hb}{\hbox{H$\beta$}}

\newcommand{\hii}{\hbox{H\,{\sc ii}}}
\newcommand{\nii}{\hbox{[N\,{\sc ii}]}}

\newcommand{\oiiitw}{\hbox{[O\,{\sc iii}]$\lambda $4363}}

\title{Energy implications of  temperature fluctuations in
photoionized plasma}
\author{Luc Binette  and Valentina Luridiana
\affil{Instituto de Astronom\ii a \\
Universidad Nacional Aut\'onoma de M\'exico}
}
\fulladdresses{

\item L. Binette: Instituto de Astronom\ii a, UNAM, Ap. 70-264, 04510 D.F.,
M\'exico (binette@astroscu.unam.mx)

\item  V. Luridiana: Instituto de Astronom\ii a, UNAM, Ap. 70-264, 04510 D.F.,
M\'exico (vale@astroscu.unam.mx)
}

\shortauthor{Binette \& Luridiana}
\shorttitle{Energy requirements of fluctuations}

\abstract{We quantify the energy radiated through all the collisionally excited
lines in a photoionized nebula which is permeated by temperature
fluctuations. We assume that these correspond to hot spots which are
the results of an unknown heating process distinct from the
photoelectric heating.  We consider all the effects of using a higher
mean temperature (as compared to the equilibrium temperature) due to
the fluctuations not only on each emission line but also on the
ionization state of the gas.  If this yet unknown process was to
radiate a fixed amount of energy, we find that the fluctuations should
correlate with metallicity $Z$ when it exceeds 0.7 solar. The
excess energy radiated in the lines as a result of the fluctuations is
found to scale proportionally to their amplitude \tsq. When referred
to the total energy absorbed through photoionization, the excess
energy is comparable in magnitude to \tsq.}
\nonstopmode

\begin{document}

%% This command is necessary to typeset the title, abstract, etc. 
\maketitle

%%
%% And here starts the text....
%%
\section{Introduction}
\label{sec:intro}
 
The temperatures of photoionized nebulae are observed to be significantly lower
when derived using  recombination lines rather than from forbidden line ratios
(e.g. Peimbert et~al. 1995).  This phenomenon has been ascribed to the
existence of temperature fluctuations permeating the nebulae. Assuming
that the  fluctuations inferred by various authors
(e.g. Peimbert et~al. 1995, Esteban et~al. 1998, Rola \& Stasi\'nska 1994) are
caused by an additional albeit {\it unknown} heating agent (beside
photoionization), we proceed to quantify the energy contribution which
this unknown heating process must contribute to the total energy
budget of the nebula in order to account for the much higher
temperatures characterizing the collisionally excited lines as
compared to those inferred from recombination lines (or nebular Balmer
continuum). In order to study the effects of arbitrary temperature
fluctuations, we first describe the modifications made to the
multipurpose photoionization-shock code \map\ (Ferruit
et~al. 1997). In   Section\,\ref{sec:cal}, we present photoionization
calculations in which we consider different levels of fluctuation
amplitudes and quantify how they alter the global energy budget
of the nebula. A brief discussion is presented in Section\,\ref{sec:con}.

\section{The energy expense caused by nebular hot spots}
\label{sec:expen}

We present a few  definitions followed by the procedure we adopt to
implement the effect of temperature fluctuations in the code \map.

\subsection{Definitions of \tsq\ and mean temperature \tmean\ }
\label{sec:def}

Following Peimbert (1967), we define the mean nebular temperature,
\tmean, as follows

\begin{equation}
\bar T_0=\frac {\int_V n_e^2 T dV}{\int_V n_e^2 dV} \; , \label{eq:to}
\end{equation}

\noindent  in the case of an homogeneous 
metallicity nebula characterized by small temperature fluctuations;
$n_e$ is the electronic density, $T$ the electronic temperature and
$V$ the the volume over which the integration is carried out. The rms
amplitude $t$ of the temperature fluctuations is given by

\begin{equation}
t^2 = \frac {\int_V n_e^2 (T - \bar T_0)^2 dV} 
{\bar T_0^2\int_V n_e^2 dV} \; . \label{eq:tsq}
\end{equation}

\noindent Note that we simplified the expression presented by
Peimbert (1967) whose definition of \tsq\ differs in principle with
each ionic species density $n_i$ while in the above equations we
implicitely consider only ionized H (by setting $n_{H^+} = n_e$).
Since \tsq\ in this paper is not an observed datum but an {\it
a~priori} global property of the nebular model, such differences are
not important.

The intensity of a recombination line is in general proportional to 
$T^{\alpha}$ while for a collisionally excited line it is proportional
to  $T^{\beta} {\rm exp}(-\Delta E/kT)$ where $\Delta E$ is the energy
separation of the two levels involved in the transition. In either
case, $\beta$ and $\alpha$ typically lie in the range $-0.5$ to $-1$.

\subsection{Approximating fluctuations as hot spots above \teq} \label{sec:hot}

Since the fluctuations' amplitudes \tsq\ derived from observations are
much larger that that predicted by photoionization models (cf.
P\'erez 1997), the solution to this inconsistency resides not in
adding an additional uniform heating/cooling term to the thermal
balance equations (like heating by dust grain photoionization) since
this would simly result in a uniform raise of $T$ and not in larger
fluctuations. What is required to increase \tsq\ in models is that
such process be operating in a {\it non-uniform} manner across the
nebula and the picture which we propose is that of many hot spots
created by this heating process.

\begin{figure}
\begin{center} \leavevmode
\includegraphics[width=0.45\textwidth]{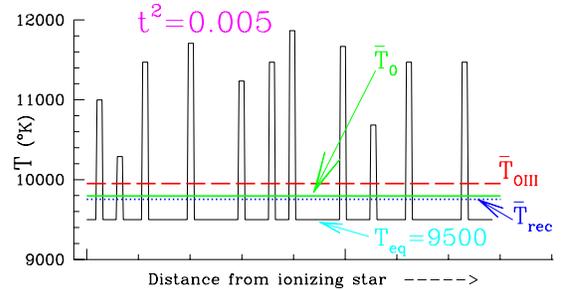}
\caption{ 
A numerical simulation of temperature fluctuations consisting of hot
spots characterized by an amplitude $\tsq=0.005$ (from
Eq.~\ref{eq:tsq}). The mean temperature \tmean\ obtained from
applying Eq.~\ref{eq:to} is 9800\,K (horizontal solid line).  Also
shown the relative position of \trec\ and $T_{[OIII]4363}$.  The
fluctuations' lower bound temperature (to be associated to \teq) is
9500\,K.
%trec and t4363 are  9753 9954
}
\label{fig:tsq} 
\end{center}
\end{figure}

It is beyond the scope of this paper to consider any particular
physical process to account for the fluctuations nor to model them in
details. Our aim is limited to study the effect of {\it ad hoc}
temperature fluctuations {\it a la} Peimbert on the energy budget
under the assumption that these arise from this unknown extra heating
mechanism acting non-uniformly across the nebula.

In photoionization calculations, it is customary to define and use at
every point in the nebula a local equilibrium temperature, \teq, which
satisfies the condition that the cooling by radiative processes equals
the heating due to the photoelectric effect. In our hot spots scheme,
by construction,  \teq\ corresponds to the temperature floor above which
take place all the fluctuations. For illustrative purposes, we show in
Fig.\,\ref{fig:tsq} a possible rendition of fluctuations (the solid
wavy line) characterized by an amplitude $\tsq=0.005$ (from
Eq.~\ref{eq:to}) and consisting of hot spots. Because \teq\ is a
minimum in the distribution of $T$ fluctuations, it defines the null
energy expense when calculating the extra energy emitted as result of
the hot spots.

\subsection{The determination of \tmean\ } \label{sec:mean}

The fluctuations are taken into account by \map\ only in the
statistical sense, that is by defining and using temperatures which
are derived from \tmean.  Given a certain amplitude of fluctuations
\tsq, we use everywhere a value for the mean temperature which is
derived from the computed local equilibrium temperature (\teq) and
which takes into account that \teq\ is a lower extremum in the
distribution of $T$ fluctuations (consisting of hot spots). To define
\tmean, we found adequate  using the following
expression (based on the inverse of Eq.\,\ref{eq:trec})

\begin{equation}
\bar T_0 \simeq \teq [1+\gamma(\gamma-1)t^2/2]^{-1/\gamma} \; . \label{eq:tmean}
\end{equation}

\noindent but where the optimum value of   $\gamma$ 
has been inferred using numerical simulations of the hot spots. One
simulation shown in Fig.\,\ref{fig:tsq} is characterized by
$\teq=9500$\,K and $\tsq =0.005$. By fitting $\gamma$ so that the value
of \tmean\ matches the numerically computed value of 9800\,K, we
obtain that $\gamma \approx -15$. Within the deduced regime in which
$\gamma \ll -1$, the above equation varies  slowly with $\gamma$.
The purpose of this expresion is that it defines consistently \tmean\
whatever the value of the equilibrium temperature calculated by
\map. Interestingly, $\gamma$ (or \tmean) is invariant to
simultaneously scaling up and down of all the hot spot amplitudes
(which is equivalent to increasing or decreasing \tsq). On the other
hand $\gamma = -15$ strickly characterizes a specific distribution of
hot spot frequencies and widths.  It is the correct value for
fluctuations resembling those depicted in Fig.\,\ref{fig:tsq} but for
other radically different distributions of hot spot widths and
frequencies, $\gamma = -15$ would only provide a first order albeit
acceptable estimate of \tmean. Despite this caveat, we estimate the
uncertainties affecting the final determination of
\gamh\ to be $< 20$\%.

\subsection{Recombination processes} \label{sec:recom}

For small fluctuations, the temperature can be expanded in a Taylor
series about the mean \tmean. In the case of recombination lines
the intensity, $I_{rec}$,  of a given line is affected by a factor

\begin{equation}
I_{rec} \propto  
\bar T_0^\alpha [1+\alpha(\alpha-1)t^2/2] \; . \label{eq:irec}
\end{equation}

\noindent This expression which can be used to compute individual
recombination line intensities in the presence of small fluctuations, is
equivalent to calculating the intensity (which is proportional to
$T^{\alpha}$) using instead the effective temperature \trec\

\begin{equation}
\trec = \langle T^\alpha\rangle^{1/\alpha}\simeq
T_0 [1+\alpha(\alpha-1)t^2/2]^{1/\alpha} \; . \label{eq:trec}
\end{equation}

As shown by Peimbert (1995 and references therein), the temperature
fluctuations have in general much {\it less} impact on recombination
than on collisional processes which are usually governed by the
exponential factor. For this reason, we adopt the simplification of
considering a single value of $\alpha = -0.83$ for all recombination
processes (such $\alpha$ is the appropriate value for the \hb\ line at
10000\,K). This approximation will allow us to use a single
temperature \trec\ when solving for the ionization balance of H, He
and all ions of metals (equations in which enter recombination rates).

If we consider the fluctuations drawn in Fig.~\ref{fig:tsq} as
example, we see that the mean recombination temperature \trec\ derived
using Eq.~\ref{eq:trec} lies slightly below \tmean.

Interestingly, in the case of the ionization-bounded dustfree models
considered in this work, the adoption of one temperature or another
when calculating recombination processes does not affect the global
energy budget of the recombination lines. In effect,
the total number of recombinations taking place across the whole
nebula must equal the number of ionizing photons produced by the UV
source, whatever the value and behaviour of the temperature. It is a
self-regulating process: a hotter nebula [which results in slower
recombinations rates and hence in a plasma containing less neutral H]
of the type discussed below will simply turn out more massive in
ionized gas in order that the number of recombinations remains equal
to the same number of ionizing photons. The sum of all hydrogen
recombination lines intensities will remain unchanged. Also constant
is the total amount of heat deposited in the nebula through photoionization.

\subsection{Collisional processes} \label{sec:coll}

To compute the forbidden line intensities, we solve for the population
of each excited state of all ions of interest assuming a system of 5
or more levels according to the ion. (In the case of intercombination,
fine structure and resonance lines, we treat those as simple 2 level
systems.)  More specifically, when evaluating the excitation ($\propto
T^{\beta_{ij}} {\rm exp}[-\Delta E_{ij}/kT] $) and deexcitation
($\propto T^{\beta_{ji}}$) rates of a given multi-level ion, each rate
$ij$ (population) or $ji$ (depopulation) is calculated using
\tmean\ (instead of \teq) and then multiplied by the appropriate 
correction factor, either

\begin{eqnarray}
{\rm cf}^{exc.}_{ij} =  
1+ \frac{t^2}{2} \Bigl[ (\beta_{ij} - 1 ) 
\Bigl(\beta_{ij} + 2 \frac{\Delta E_{ij}}{k \tmean} \Bigr) +
\Bigl(\frac{\Delta E_{ij}}{k \tmean}\Bigr)^2
\Bigr]  \label{eq:exc}
\end{eqnarray}
\noindent in the case of excitation, or
\begin{eqnarray}
{\rm cf}^{deexc.}_{ji} = 
 1 +  \beta_{ji} (\beta_{ji} - 1 ) 
\frac{t^2}{2} \; , \label{eq:deexc}
\end{eqnarray}
\noindent in the case of deexcitation.
These factors result in general in an enhancement of the collisional
rates in the presence of temperature inhomogeneities. They are adapted
from the work of Peimbert et~al. (1995) and were applied to all
collisionally excited transitions.

For the case of the particular rendition of the fluctuations depicted
in Fig.\,\ref{fig:tsq}, the mean {\it collisional} temperature
characterizing the \oiiitw\ line (Peimbert et~al. 1995) is shown by the 
horizontal dash line. Despite the large $\Delta E_{ij}$ involved in
the emission of the line, the distance between this temperature and
\tmean\ is nevertheless smaller than that separating \tmean\ from
\teq. Therefore, in some instances the extra energy radiated through
hot spots might depend as much on the distance separating \tmean\ from
\teq\ than on the above correction factors.

Inspection of the line ratios calculated with \map\ using \tmean\ and
the above correction factors confirms (as expected) that the higher is
$\Delta E_{ij}$ the higher the line intensity enhancement (at constant
\tsq).  It can be shown on the other hand that the far infrared lines
(or any transition for which ${\rm exp}[-\Delta E_{ij}/kT] \approx 1$)
are less affected by the fluctuations (similarly to the recombination
lines) and can even become weaker as a result of the fluctuations, in
contrast to most collisionally excited lines.

\subsection{The energy radiated through hot spots} \label{sec:ener}

Our aim is to quantify the excess energy generated by temperature
fluctuations under the assumption that these are caused by a putative
heating mechanism which operates within small regions randomly
distributed across the nebula. To calculate this energy we simply
integrate over the nebular volume $V$ the luminosity of each line (or
transition) $ij$ using the statistically determined local \tmean\ and
the multiplicative correction factors of Eqs~\ref{eq:exc} and
\ref{eq:deexc}, and then subtract the corresponding luminosity
obtained by using the equilibrium temperature \teq\ instead. This
excess energy radiated in the form of collisionally excited lines can
be normalized respective to the total photoheating energy
available. This defines the quantity \gamh\

\begin{eqnarray}
\gamh\ & = &
 { \sum_{ij} \int_{V} [4\pi j_{ij}^{fluc} -  4\pi
j_{ij}^{eq}] \; dV} \over
{\sum_{ij} \int_{V}   4\pi j_{ij}^{eq} \; dV + \sum_{k} \int_{V} q_k^{eq} \; dV  }   \nonumber \\ & & \\
     & = & {{L_{fluc} - L_{eq}}\over{L_{eq} + Q_{eq}}} \;,  \label{eq:gam}
\end{eqnarray}

\noindent where $j_{ij}^{fluc}$ corresponds to
the local nebular emissivity of line $ij$ calculated using \tmean\ and
taking into account the above correction factors while $j_{ij}^{eq}$
is the corresponding emissivity assuming equilibrium temperature
everywhere. The term with $q_k^{eq}$ corresponds to various cooling
rates from processes {\it not} involving line emission such as
freefree emission, while $Q_{eq}$ represents the total volume integrated
value of this term.  $L_{fluc}$ and $L_{eq}$ correspond to the
integrated energy loss across the whole nebula due to collisionally
excited lines with fluctuations and without, respectively. Since
\teq\ must satisfy the condition that the cooling rate equals
everywhere the heating rate, we obtain that $L_{eq} + Q_{eq}$ is also
equal to the total energy deposited into the electronic gas by the
photoelectric effect. 

A larger fraction of the ionizing radiation simply keeps the
nebula ionized (resulting in recombination lines) but does not affect
the nebular temperature. An alternative way therefore of expressing the
importance of the excess cooling due to the fluctuations is to use as
reference the {\it total} energy absorbed by the nebula {\it
including} the energy emitted as recombination lines and nebular
recombination continuum. We define \gama\ as follows

\begin{equation}
\gama\ = {{L_{fluc} - L_{eq}}\over{L_{total}^H}} \; ,  \label{eq:heat}
\end{equation}

\noindent with $L_{total}^H$  the ionizing luminosity of the exciting star

\begin{equation}
L_{total}^H = \int_{\nu_0}^\infty {{L_\nu}} \,d\nu \; ,  \label{eq:ener}
\end{equation}
\noindent  where $L_{\nu}$ is the  energy luminosity distribution of
the ionizing star and $\nu_0$ the frequency corresponding to the
ionization treshold of H. To be consistent with this definition, we
must consider only nebular models which are ionization-bounded and
fully covering the ionizing source (over 4$\pi$ sterad).  Depending on
the hardness of the ionizing radiation, \gama\ turns out to be 2--3
times smaller than \gamh\ because of the larger fraction of the absorbed
energy which goes into photoionizing rather than into heating the gas.

In summary, the modifications made to \map\ to consider the effects of
$T$ fluctuations not only include the calculation of the line
intensities using the formalism described in Peimbert et~al (1995) but
also considers their impact on the ionization balance across the
nebula through the use of \trec\ (instead of \teq) to derive the
recombination rates. Since the nebula
is substantially hotter {\it on average} when $\tsq > 0$, it will be more
ionized since the recombination rates are slower at higher
temperature. This in turn results in a lower photoionization and
heating rates given that there is less neutral H in the nebula,
therefore the equilibrium temperature \teq\ computed by \map\ will be
lower than without fluctuations. All these effects have been taken
into account self consistently and do {\it not} affect the energy
conservation principle in the case of ionization-bounded nebulae as
discussed above and in in Sect.~\ref{sec:recom}.

\section{Model calculations} \label{sec:cal}

We have explored the behaviour of \gamh\ in photoionization models of
different metallicity ($Z$), excitation (\up) and different spectral
energy distributions (hereafter SED). We will express the nebular
metallicity with respect to the solar abundances (from Anders \&
Grevesse 1989) for which we take that $Z=1$. To define other
metallicities, we simply scale the abundances of all the metals
respective to H by a constant multiplicative factor equal to $Z$. We
summarize below our results under various model conditions.

For the photoionized \hii\ regions, we have selected unblanketed LTE
atmosphere models from Hummer \& Mihalas (1970) of temperatures \teff\
of 40000\,K, 45000\,K and 50000\,K [see Evans (1991) for a comparative
study of nebular models using different model atmospheres]. To
represent planetary nebulae, we simply employed black bodies of
$10^5\,$K and $10^{5.3}\,$K truncated at 54.4\,eV.  The geometry
adopted in the calculation is plane-parallel  with a gas
density of $n_H = 10\,\cmc$ in all cases. The excitation of the nebula
is defined by the excitation parameter \up\ as follows

 \begin{equation}
U = {{1}\over{cn_H}} \int_{\nu_0}^\infty {{L_\nu}\over {4 \pi r^2 \, h\nu}}
\,d\nu  = {{\varphi_H}\over {c n_H}} \; ,  \label{eq:upar}
\end{equation}
\noindent  where $c$ is the speed of light, $h$ the Planck constant
and $r$ the distance of the slab from the ionizing star. \up\ is the
ratio between the density of ionizing photons impinging on the slab 
($\varphi_H/c$) and the total H density. All the calculations  carried out
were ionization-bounded.

\begin{figure}
\begin{center} \leavevmode
\includegraphics[width=0.95\columnwidth]{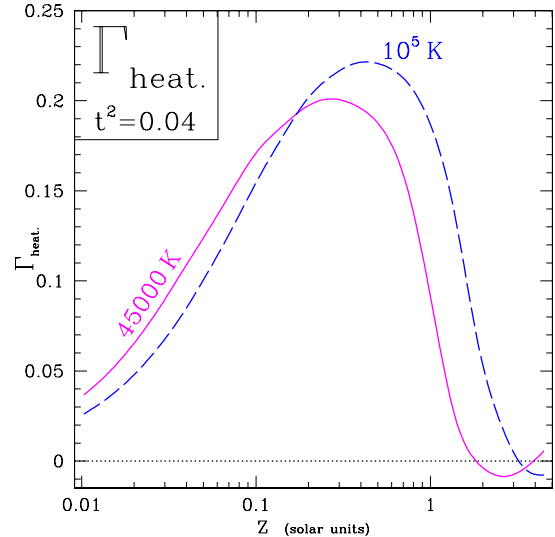}
\caption{ 
Behaviour of \gamh\ in a sequence of photoionization models which have
different nebular metallicities (relative to solar $Z=1$). Two SED
used in the calculations consisted of a 45000\,K and a 100000\,K star,
respectively. In all models $\tsq = 0.04$ .}
\label{fig:zseq} 
\end{center}
\end{figure}

Since hotter SEDs result on the other hand in much higher
photoelectron energies and hence hotter nebulae at all metallicities,
the maximum of the long-dashed curve corresponding to the $10^5$\,K
blackbody in Fig.\,\ref{fig:zseq} is shifted towards higher $Z$
relative to the 45000\,K SED.

\subsection{Dependence of excess heating on metallicity}

We have calculated nebular models of different metallicities covering
the range 1\% solar ($Z=0.01$) to 4.7 times solar. In the models
presented in Fig.\,\ref{fig:zseq} all other parameters are identical,
namely, \up\ = 0.01, \tsq\ = 0.04 and a spectral energy distribution
(SED) having either \teff\ = 45000\,K or 100000\,K.  It can be seen
that a maximum in \gamh\ occurs within the range $Z \sim $ 0.2--0.4.
For the \hii\ region models, the average values for \teq\ across the
nebulae are $\sim 15000$\,K, 9000\,K, 5000\,K and 1500\,K for the
$Z=0.01$, 0.7, 2.5 and 4.7 models, respectively. Given that the
nebular temperatures decrease monotonically with increasing $Z$, the
curves' behaviour can be understood as follows: at very low
metallicities $Z \ll 0.2$, the forbidden lines of metals are not the
main cooling agent and the fluctuations have therefore a negligible
impact on the total cooling. At higher $Z$ values around solar,
however, the cooling due to metals is very large and the nebula turns
out much cooler, to the extent that many optical lines become now less
intense despite the increase of the metal abundances. At even higher
$Z$, the optical lines are progressively `switched off' and cannot
contribute to the cooling of the nebula.  In this regime, these lines
have $\Delta E \ll k\tmean$ and become somewhat brighter as the
temperature is further lowered, explaining why the fluctuations now
cause \gamh\ to become negative. The final rise at the upper $Z$ end
in the \hii\ region model sequence (solid line) reflects the fact that
at such low temperature the cooling can only be carried out by the low
energy transitions (far infrared lines) where again $\Delta E > k\tmean$.

\begin{figure}
\begin{center} \leavevmode
\includegraphics[width=0.95\columnwidth]{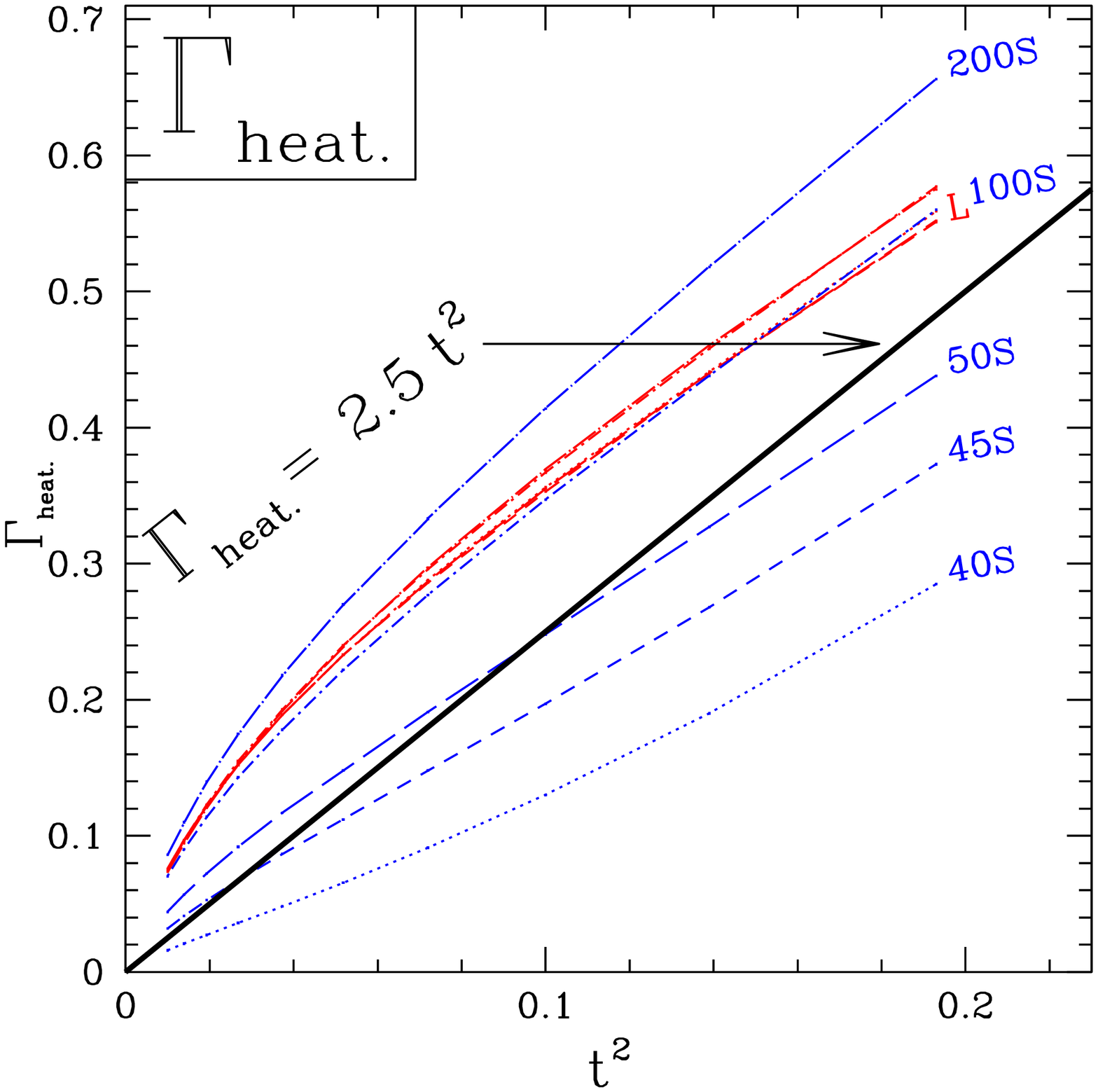}
\caption{ 
Behaviour of \gamh\ with increasing \tsq. Two metallicities were used:
solar (labelled S) and 0.2 solar (labelled L) and 4 SED:
40000\,K,45000\,K, 50000\,K, 100000\,K and 200000\,K labelled 40, 45,
50, 100 and 200, respectively (the L sequences are too close to allow
labelling of each SED).  \gamh\ becomes  approximately
linear at larger \tsq. For comparison we show a thick line
representing $\gamh = 2.5 \tsq$.}
\label{fig:tsqh} 
\end{center}
%\end{figure}

%\begin{figure}
\begin{center} \leavevmode
\includegraphics[width=0.95\columnwidth]{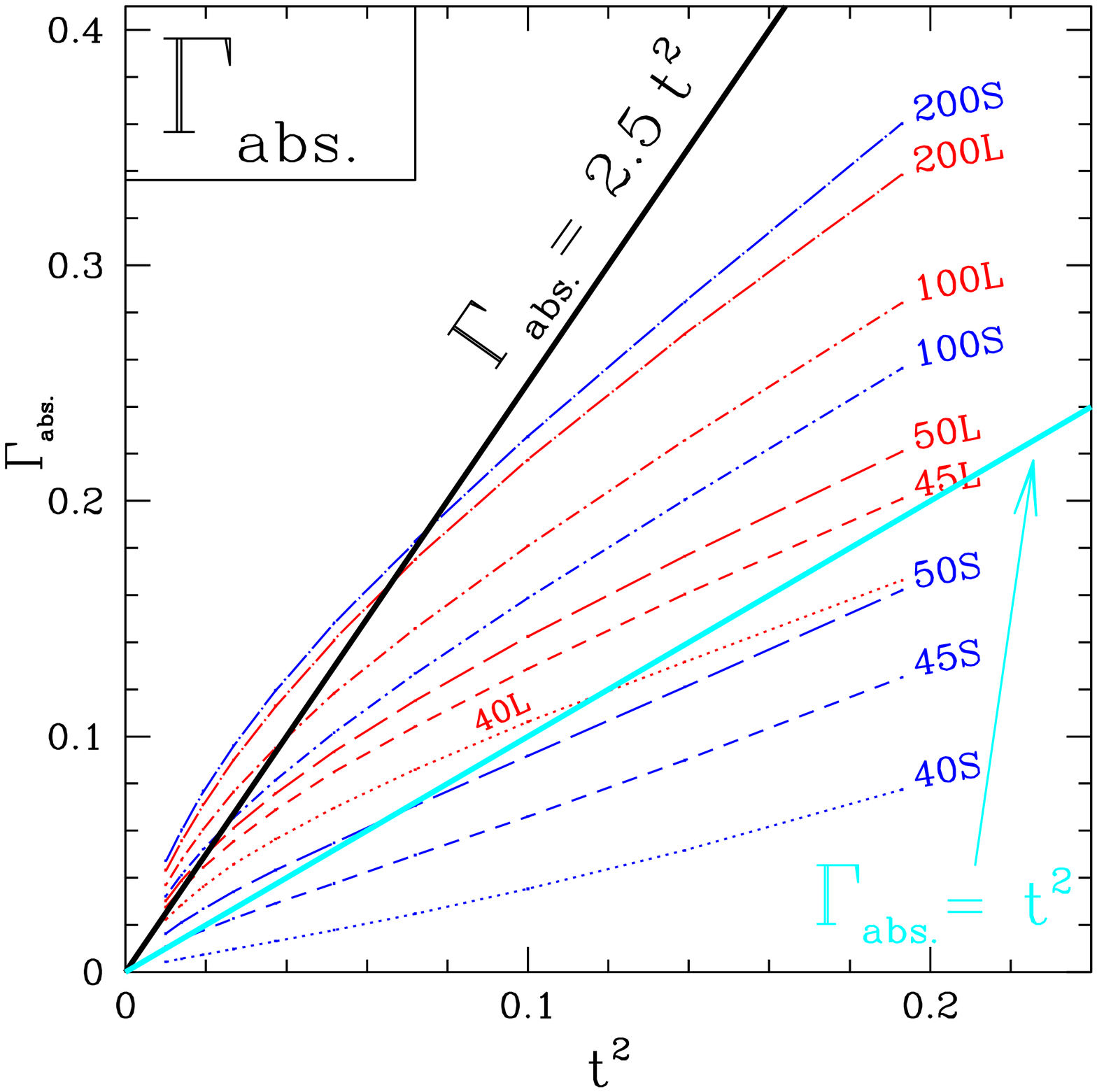}
\caption{ 
Behaviour of \gama\ with increasing \tsq. Same notation as in
Fig.~\ref{fig:tsqh}.  For comparison we show  thick lines representing
$\gama = \tsq$ and $2.5 \tsq$, respectively.}
\label{fig:tsqa} 
\end{center}
\end{figure}

The abrupt decrease of \gamh\ above solar metallicity has interesting
consequences if the unknown heating process responsible for the
fluctuations needed to radiate a comparable amount of energy in
different objects. This could arguably be the case for instance if
this process was reconnection of magnetic field lines. All other
factors being equal, nebulae three times solar would require much
larger amplitude turbulences (larger \tsq) to radiate the same amount
of energy than a solar metallicity nebula. In this case we might
expect to observe a correlation between \tsq\ and metallicity (beyond
$Z \ge 0.7$).

\subsection{Dependence of excess heating on \tsq\ }

In the following calculations, we adopt two representative metallicities
of $Z=0.2$ and $Z=1$ (solar).  We found no clear trends across
different SEDs of how \gamh\ varied with \up\ and  therefore we only report
results concerning a single ionization parameter of value $10^{-2}$.
Campbell (1988) has shown that the range of \up\ for most \hii\
galaxies lies in the range $10^{-2.6}$ to $10^{-1.8}$.

In Figs.\,\ref{fig:tsqh} and \ref{fig:tsqa}, we show the behavior of both
\gamh\ and \gama\ as a function of increasing \tsq\ of the models. Each 
line corresponds to a given SED and $Z$. The increase in \gamh\ is
steeper at small values of $\tsq \la 0.1$ followed by a more linear
regime with a slope $\le 2.5$ at large \tsq.  The radiated energy
contribution from the fluctuations (compared to photoelectric heating)
is of order $2.5 \tsq$ (Fig.~\ref{fig:tsqh}) but with a wide
dispersion when $Z=1$ (curves labelled S). For $Z=0.2$ there is no dependence
on the SED and the curves almost superimpose each other.

If we consider the total energy budget (photoheating plus
recombination energy, see Eq.~\ref{eq:heat}) and not just the
photoelectric heating part, we find that \gama\ is comparable in
magnitude to \tsq\ (Fig.~\ref{fig:tsqa}) but again with a wide
dispersion which result from differences in either $Z$ or the SED.

\section{Discussion} \label{sec:con}

If the turbulences are the result of heating by a yet undiscovered
process, we infer from Fig.~\ref{fig:tsqh} and \ref{fig:tsqa} that the
extra energy radiated via {\it all} the collisionally excited lines
due to the turbulences is a substantial fraction ($\gama >10$\%) of
the total photoionization energy whenever $\tsq\ \ga 0.02$ for
planetary nebulae and $\tsq\ \ga 0.08$ for \hii\ regions,
respectively.  Taking instead as reference only the heating by
photoionization (\gama), these limits are reduced by two. In this case
almost all nebulae would radiate more than 10\% (\gama) of their
energy as a result of $T$ fluctuations. In effect, typical empirically
determined \tsq\ values for galactic \hii\ regions lie in the range
0.02--0.06 (Luridiana 1999). For extragalactic \hii\ regions
(Luridiana et~al. 1999, Gonz\'alez-Delgado et~al. 1994) and planetary
nebulae (Peimbert et~al. 1995), even larger values have been
encountered, of order 0.1 or more\footnote{Measured values of
\tsq\ in excess of 0.1 should be considered only as rough estimates
since the assumed regime of small fluctuations does not apply
anymore.}. In those cases, the energy envolved can be a large
fraction of the energy budget especially if the exciting stellar
temperature exceeds $10^5\,$K.

The underlying assumption behind our calculations is that an external
heating agent is at work to generate $T$ fluctuations. We should point
out that there exist alternative explanations to the fluctuations
which rest on photoionization alone and have not been completely ruled
out. Possible mechanisms which we plan to study in some details:

\begin{enumerate}
\item{\it Metallicity inhomogeneities}
Temperature fluctuations would result naturally from nebulae
consisting of a multitude of condensations of greatly varying
metallicities (see Torres-Peimbert, Peimbert \& Pe\~na 1990)
\item{\it Gas expansion}
Fluctuations could arise from an outflowing wind generated at the
surface of dense condensations (problyds) as a result of the champagne
effect.  Rapid adiabatic expansion would result in overcooled emission
regions in the wake of the wind (resulting in negative $T$ fluctuations
with respect to \teq).
\item{\it Ionizing source variability}
A rapidly varying ionizing field will generate a sequence of outwardly
propagating ionization and recombination fronts. The basic asymmetry
existing between ionization fronts (propagating at a speed limited by
changes in opacity) and recombination fronts (non-propagating), would
lead to temperature fluctuations.
\end{enumerate}

To our knowledge, none of these mechanisms have yet been incorporated
explicitly in any nebular model and we therefore cannot assess their
particular merits. Mechanisms which are good candidates to explain the
extra heating assumed in this work (see also discussion by Stasi\'nska
1998) and which are not directly related to photoionization, have been
proposed and include the following processes:
\begin{enumerate}
\item{\it Shock heating} 
Either as a result of stellar wind or from supernovae. In the case of
giant \hii\ regions, this later mechanism was   favored by
Luridiana et~al. (1999).
\item{\it Magnetic heating} Turbulent dissipation from Alf\'en waves
and magnetic line reconnection has been proposed to explain the
variations of the \nii/\ha\ ratio emitted by the warm diffuse ionized
gas in the Galaxy (Reynolds et~al. 2000). In the case of planetaries,
the large values of \tsq\ found would then make sense in the light of the
success of the magnetically accelerated wind models of Garc\'\i
a-Segura (2000).  Furthermore, Peimbert et~al. (1995) has shown that
the nebulae with the highest gas velocity dispersion show  the highest
values of \tsq, which would be consistent with an increasing
role played by magnetic acceleration of the gas in the objects with
the largest velocity dispersion.
\end{enumerate}

In summary, the current work shows how \gama\ and \gamh\ vary with
metallicity and \tsq\ as well as how these quantities are affected by
the ionizing energy distribution. Our results can be translated into
energy requirements (function of \tsq) which  competing
explanations of the temperature fluctuations must satisfy and, hence,
can be used to probe their respective viability.

\acknowledgements The work of LB was supported by the CONACyT grant 32139-E.

%\adjustlastcols

%% When using the rmaacite package, the \bibitem command should be of
%% the format: 
%%
%%             \bibitem[AUTHOR<YEAR>]{KEY} 
%%
%% so that the \cite{KEY}, etc. commands will work properly. 
%% 
%% If you are doing the citations manually, then you can just use
%% `\bibitem{}' instead. This will give you a warning about
%% `multiply-defined labels' which you can safely ignore.
%% 

\end{document}